The precise determination of mass through the oscillations of a very high-$Q$ superconductor oscillating system.

Osvaldo F. Schilling
Departamento de Física, Universidade Federal de Santa Catarina, Campus, Trindade, 88040-900, Florianópolis, SC. Brazil.
Email: osvaldo.neto@ufsc.br

Abstract:
The present paper is based upon the fact that if an object is part of a highly stable oscillating system, it is possible to obtain an extremely precise measure for its mass in terms of the energy trapped in this resonance. The subject is timely since there is great interest in Metrology on the establishment of a new electronic standard for the kilogram. Our contribution to such effort includes both the proposal of an alternative definition for mass in terms of energy, as well as the description of a realistic experimental system in which this definition might actually be applied. The setup consists of an oscillating type-II superconducting loop ( the SEO system) subjected to the gravity and magnetic fields. The system is shown to be able to reach a dynamic equilibrium by trapping energy up to the point it levitates against the surrounding magnetic and gravitational fields, behaving as an extremely high-$Q$ spring-load system. The proposed energy-mass equation applied to the electromechanical oscillating system eventually produces a new experimental relation between mass and standardized constants.

1. Introduction..

Mass is traditionally interpreted as the inertia content of a body[1]. The greater the parameter named mass($m$), the more difficult it is for a given force $F$ to produce a change in the objects state of motion. Newton's Second Law describes this effect as $F = m\,a$, which serves as a formal definition for mass. However, the parameter mass is part of other dynamic quantities like momentum and energy. The measurement of such quantities in place of force and acceleration can be advantageous in both experimental and theoretical grounds, in such a way that it may become convenient to adopt an alternative definition for mass based upon other dynamic quantities. In the present paper we describe how the establishment of a particular mathematical relation between mass and energy can lead to a neat *new* magneto-dynamic definition for the mass of an object.

We shall be interested in systems consisting of a macroscopic object in some way attached to an environment or device capable of producing over it some restitution force. That is, if the object is initially at rest and is suddenly subjected to an external force there will be a reaction against such force tending to restore the object to its initial location. We will specialize in the well known Basic Physics example of one such system, which is the spring-load set. We develop the theory in full detail to facilitate the comprehension of the comparisons to be made with the electromechanical system to be discussed in sections 2 and 3. The spring-load system will be taken initially at rest with no stored energy, until on time $t=0$ a fixed external force $F$ is imposed upon it and kept constant thereafter. Newton's Law describes the displacement $x(t)$ of the spring and load as:

$$mx''(t) = F - k\,x(t) \qquad (1)$$

Here $k$ is the elastic constant of the spring, and $x''(t)$ denotes the second time-derivative of the displacement. The initial conditions are $x(0) = x'(0) = 0$. The solution of (1) predicts a harmonic oscillation of the load around a position displaced of $F/k$ from its initial position, with the amplitude of oscillation given by

$$\Delta x(t) = -(F/k)\cos(\omega t) \qquad (2)$$

with $\omega = (k/m)^{1/2}$ the frequency of the oscillations. Note that we are ideally assuming that no damping occurs, so that the total energy given by

$$E = 2F\,|(\Delta x)_{max}| \qquad (3)$$

is conserved. Eq.(3) states that the total energy contained in the system is given by the work done upon the system by the force $F$ along the range of oscillation. By taking the time derivative of (2) we obtain the velocity of the oscillating load, and realize that (3) can be rewritten in the concise form

$$m = E/(2\,v_m^2) \qquad (4)$$

Here $v_m$ is the maximum velocity of the load, given by $F/(k\,m)^{1/2}$. We note the following. Firstly, the form of eq. (4) independs on the particular expression for $F$, as well as it independs on the details of the restitution force, which is simply taken as linear in the displacement. Provided the system is initially fully at rest the imposition of any constant force at $t=0$ will result in (4). Secondly, this definition of mass takes into account the forces the object is subject to, and should be valid provided full equilibrium with these force fields is attained, i.e., by reaching a resonating state.

The applicability of this definition course depends upon the ability of the system in conserving energy and upon the possibility of measuring $E$. Therefore, we have shown that mass can be defined in terms of energy and velocity, a definition that can be very convenient provided a suitable system is devised to explore it.

If damping is negligible the system oscillates in a state of dynamic equilibrium with all the force fields, internal or external, around it. We will be interested in the particular case of $F = mg$, the weight of the load oscillating in the vertical direction. In this case $v_m = g/\omega$, and the oscillations amplitude is $g/\omega^2$.

Equation (4) can be useful provided energy losses are vanishingly small. However, it is well known that real spring-load systems are highly dissipative[2]. It is possible to define the quality factor $Q$ for an oscillating system by the product of the frequency of oscillations times the total stored energy divided by the dissipated power. Therefore, the remainder of this paper will deal with the description of a system which by design would display a $Q$ of $10^7$ or more, so that the application of (4) is justified. The present investigation is motivated by a particular (long-standing) Metrology problem, namely the determination of a new international standard for the kilogram in terms of an experimental expression involving well defined standardized parameters like the electronic charge and Planck's constant. The current stage of such efforts has been described for instance in [3,4]. Our objective hereafter is to show that a relatively simple system, in a single experiment, and adopting eq. (4) as the definition of mass might be able to produce an extremely precise measure for the mass of an object. The main result of this work is a new expression relating mass to other constants of Nature to be obtained by experiment (section 2), followed in section 3 by a full discussion of the principles behind the design of the experimental setup. Section 4 lists the Conclusions.

2. The superconductor electromechanical oscillator( SEO).

In order to obtain a real application for (4) we need to find a system in which losses are practically eliminated. Air friction can be virtually eliminated by working under high vacuum and at very low temperatures, to markedly decrease the atmosphere viscosity. Heat dissipation in the restitution process responsible for the oscillations is much more difficult to eliminate. However, as we show hereafter it has been possible to design a system in which all dissipative processes can be diminished to the point that the application of (4) becomes realistic. This system has been devised and previously analyzed in detail by Schilling[5,6] (although its equations of motion had been independently published much earlier by Saslow[7]), and consists of a type-II superconducting rectangular loop of mass $m$, subjected to magnetic fields and gravity. In this section we present the main results obtained from the analysis of the loop motion, including the proposal of a new relation between the loop mass and the total energy which follows from (4).

The superconductor electromechanical oscillator(SEO) is shown in Figure 1.The tridimensional equivalent of this system, comprising three rectangular oscillating loops has recently been discussed in [8] and used as a ressonator and part of a quantum magneto-mechanical device. In Fig.1 the loop is hanging upright under the effect of gravity, with magnetic fields $B_1$ and $B_2$ normal to its surface. The need for two fields is explained in section 3. We will assume initial conditions identical to those for the spring-load system of section 1, that is, the loop at rest with no currents flowing and thus no stored energy. The fact that the loop is superconducting in principle suggests that there will be no heat losses, but as we describe in section 3 several conditions must be met in order that dissipative effects actually become negligible. For the moment let us assume that such steps have been taken, and no dissipation occurs if electrical currents flow around the loop. A consequence of such null electrical resistivity is that no electromotive force is developed around the loop, associated with current transport. When the loop is released and falls under the effect of gravity, the variation in the applied magnetic flux across its open area $\Phi_m$ will be compensated by flux ( $L\,i$ ) generated by supercurrents $i(t)$ around its perimeter, in such way that the application of Faraday's Induction Law gives a null electomotive force $\varepsilon$:

$$\varepsilon = 0 = -d\Phi_m/dt - L\,di/dt \qquad (5)$$

Here $L$ is the self-inductance of the loop. This is an example of the property of flux conservation displayed by superconducting loops and rings. The loop will move with speed $v$ described by Newton´s Law:

$$m\,v'(t) = mg - iaB_0 \qquad (6)$$

In equation (6) we account for the opposing magnetic forces the fields $B_1$ and $B_2$ impose upon the currents in the two horizontal sides of the loop. In Figure 1, $a$ is the length of the lower horizontal side of the loop subjected to $B_2$. We define $B_0 = B_1 - B_2$.

From Figure 1, $d\Phi_m/dt = -B_0 a v$. Thus, from (5) one obtains a relation between $v$ and $di/dt$. Taking the time derivative of (6) and eliminating $di/dt$ from (5) one obtains:

$$v''(t) + (B_0^2 a^2/mL)\,v(t) = 0 \qquad (7)$$

which is the differential equation obeyed by the velocity of a harmonic oscillator. Assuming zero initial speed and an initial acceleration $g$, eq. (7) can be solved:

$$v(t) = (g/\Omega) \sin(\Omega t) \qquad (8)$$

From (7), the natural frequency of the oscillations is $\Omega = B_0 a/(mL)^{1/2}$. As the loop is released from rest, the assumed perfect flux and energy conservations will make this initial position the uppermost point of its oscillating path ( measured from the middle point of its oscillations range), with $x(0) = -g/\Omega^2$. The position is described by the equation

$$x(t) = -(g/\Omega^2)\cos(\Omega t) \qquad (9)$$

which is equivalent to (2) if $F = mg$. The amplitude of the oscillating motion is $x_0 = g/\Omega^2$. It is clear that such equations are equivalent to the ones in section 1 for the spring-load system, with $\omega$ replaced by $\Omega$. It is possible then to combine (5)–(6) to obtain an equation for the current $i(t)$:

$$(B_0 a/\Omega^2)\, d^2 i/dt^2 = mg - B_0 a i \qquad (10)$$

whose solution is

$$i(t) = (mg/(B_0 a))(1 - \cos(\Omega t)) \qquad (11)$$

for $i(0) = di/dt(0) = 0$. We define $i_0 \equiv mg/(B_0 a)$.

It is clear that (3) also applies here, so that the same energy conservation $E = 2m v_m^2$ is obeyed, with $v_m = g/\Omega$ from (8), as in section 1. We define $\Phi = L i_0$ as the flux threading the loop due to the average loop current value $i_0$, resulting after a simple manipulation in $E = 2\Phi^2/L$.

These two expressions for $E$ lead to the following definition for the mass of the loop:

$$m = \Phi^2/(L v_m^2) \qquad (12)$$

This is the main result of this work. A superconducting quantum interference device (SQUID) with a proper gradiometer as proble, with the oscillating loop between its coils, will measure the flux $\Phi$ due to the average current in the loop within a fraction of the flux quantum $\Phi_0 = h/(2e)$, with $h$ the Planck constant and $e$ the electronic charge. Optic-electronic measurements will accurately give the speed, and $L$ is a parameter for the loop ( to be measured in advance in a separate experiment). Therefore, a simple experimental relation between mass and standardized constants is obtained from this experiment, provided $Q$ is very large. The entire setup and theoretical analysis is considerably simpler than the Watt Balance Method adopted in recent attempts for the determination of the new standard for the kilogram[3,4].

In the next section the actual design and implementation conditions necessary for this experiment are described in detail.

3. Implementation of the SEO experiment.

The design of the system has been discussed in [5,6] but we describe here the main details. This discussion does not includes further details on the experimental techniques to be adopted, which should be considered in due course. The design should start by the choice of a material to make the loop. Superconductors exist in the so-called types I and II[9]. The problem with type I materials is that magnetic levitation requires currents in the wire that would probably be too high for a type-I material to sustain, and the wire would become resistive. The application of *B* might also turn parts of the wire into the "intermediate state", in which there are normal( resistive) regions mingled with superconducting regions, which is not desired here. By employing type-II materials, the mentioned difficulties with the magnitude of the currents and the intermediate state can be circumvented, since they are suitable for high currents and fields. However, type-II materials have problems of their own. Flux penetrates in type-II materials in the form of flux-lines(FLs) for fields greater than the lower critical field[9]. Figure 1 shows two magnetic fields $B_1$ and $B_2$ greater than the lower critical field in order to actually impose the full type-II superconducting conditions to the whole length of wire, avoiding the intermediate state of lower fields. The FLs carry a quantum of magnetic flux each and their cores contain electrons in the normal state. If these lines are not properly pinned, their motion under magnetic forces will generate heat. Fortunately, suitable working conditions can be devised to drammatically decrease such effects[5]. Commercial type-II superconductors are materials like Nb-Ti alloys. These materials contain nanometric-scale Ti inclusions that interact quite strongly with the FLs and pin then down. However, the motion of the loop will generate oscillating currents quite close to the wire surface[5], and these currents will produce a very small field that might disturb the FLs in an irreversible, hysteretic way, which must be avoided. If such fields are kept below a threshold value[5,10-11] there will be no major FLs displacements. The FLs will remain pinned, only slightly displaced from their equilibrium positions in the absence of that ripple field. The losses will be negligible for all practical effects, as calculated for a case study in [5].

Although we can conclude that losses within the wire can be made very small indeed, a remaining important source of energy loss is the drag of the loop against the cryostat atmosphere, even under high-vacuum. Temperature should be kept at the lowest possible level, preferably below 0.1K. It has been shown in [5] that if the above conditions alone dominate the energy dissipation process a quality factor *Q* on the order of $10^{10}$ might be reached. However, other possible sources of energy dissipation should be avoided. Any electric or magnetic coupling of the loop with its surroundings will incur in extra losses. Eddy currents might be induced in metallic parts like the cryostat itself and the magnets that produce the field. Electrically insulating materials should therefore be used. In addition, the loop will produce an external oscillating field on the order of some gauss, and if permanent magnets are adopted the material used in the magnets should be hard enough not to display hysteresis under this kind of field. If a superconducting magnet is used to produce the field imposed to the loop the mutual inductance effect would be part of the solution of the problem and there would be no major dissipative terms added.

It is possible to relax the requirement of ultra-low temperatures, and work at, say, 4.2K and high-vacuum. The viscosity of the atmosphere will increase the drag effect upon the loop by a fator of about 1000[5,12]. The theoretical quality factor will drop from $10^{10}$ to the $10^7$ range, which is comparable to that for electronic clocks used commercially, but even so the system will still be very close to ideal conditions. Experiments at 77K might even be tried, since the drag effect will increase by a factor of about 50 between 4.2 and 77K, which might still be acceptable. In this case the loop might be made of a high-temperature superconductor, and the whole setup would require simpler refrigeration techniques.

4. Conclusion

This paper has discussed in detail how an alternative definition of mass can be proposed for a very high-$Q$ oscillating spring-load system. Provided the system starts entirely at rest, the application of a constant force at $t=0+$ will lead to a 3-parameter relation between total energy, mass and the square of the oscillations speed. Such relation, eq. (4), is the expression of the resonating equilibrium state achieved by the load in the presence of the surrounding force fields. The concept is applied to the extremely high-$Q$ SEO, which is a system that traps energy from the surrounding magnetic and gravitational fields and levitates in a nonhysteretic motion. The entirely new equation (12) relates the mass of the loop with standardized constants. Such experimental system should be of great interest for Metrologists since it provides an essentially magneto-optical technique for obtaining an extremely precise measure of the mass of an object .

Figure Caption:
Figure 1
A type-II superconducting loop levitates under the action of two magnetic fields $B_1$ and $B_2$ greater than its lower critical field ($B_1 > B_2$ in the figure and the fields point into the page). $F_1$ and $F_2$ are the opposing Lorentz forces that the fields impose upon the clockwise current ( dashed line) induced in the loop when it is initially released and moves under the effect of gravity **g**. A resonating state is achieved, with the loop reaching dynamic equilibrium with the force fields around it.

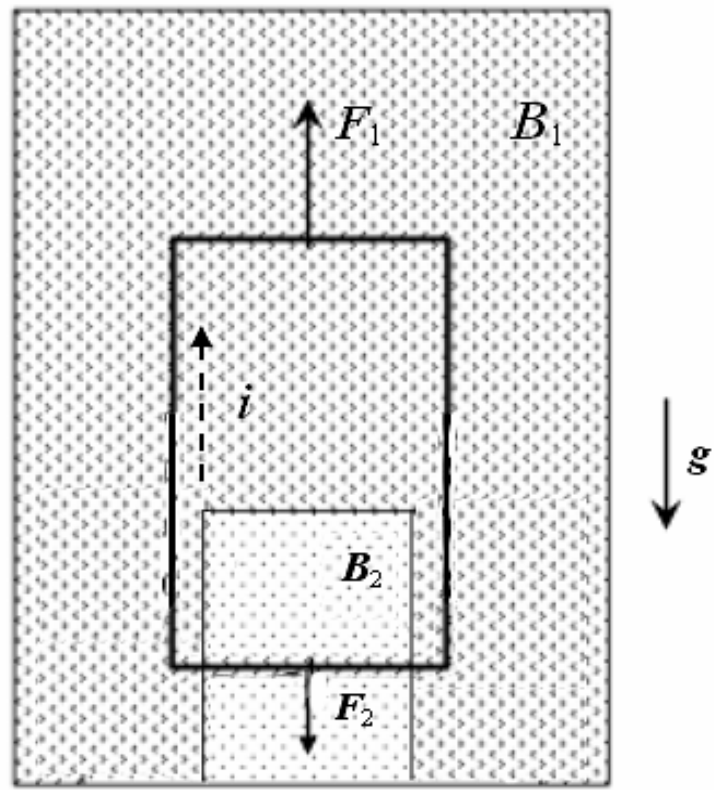